\pdfoutput=1
\documentclass[a4paper,12pt,abstract=true]{scrartcl}
\usepackage{authblk}
\usepackage{amsmath,amssymb}
\usepackage{bm}
\usepackage{graphicx,color}
\usepackage[hidelinks]{hyperref}

\usepackage[all,warning]{onlyamsmath}
\RequirePackage[l2tabu, orthodox]{nag}

\newbox{\ORCIDicon}
\sbox{\ORCIDicon}{\large
                  \includegraphics[width=0.8em]{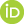}}

\begin{document}

\titlehead{\hfill OU-HET-1088}

\title{\LARGE Alternative method of generating gamma rays 
       with orbital angular momentum}

\author[1]{Minoru~Tanaka\,\href{https://orcid.org/0000-0001-8190-2863}
                               {\usebox{\ORCIDicon}}
           \thanks{Email: \texttt{tanaka@phys.sci.osaka-u.ac.jp}}}

\affil[1]{Department of Physics, Graduate School of Science, 
          Osaka University, Toyonaka, Osaka 560-0043, Japan} 

\author[2]{Noboru~Sasao
           \thanks{Email: \texttt{sasao@okayama-u.ac.jp}}}
\affil[2]{Research Institute for Interdisciplinary Science, 
          Okayama University, Okayama, 700-8530, Japan}

\date{\normalsize\today}

\maketitle

\begin{abstract}
We study an alternative method of generating gamma rays 
with orbital angular momentum (OAM). 
Accelerated partially-stripped ions are used as an energy 
up-converter.
Irradiating an optical laser beam with OAM on ultrarelativistic ions,
they are excited to a state of large angular momentum.
Gamma rays with OAM are emitted in their deexcitation process. 
We examine the excitation cross section and deexcitation rate.
\end{abstract}

\section{\label{Sec:Intro}Introduction}
It is now well known that light (electro-magnetic wave) has angular
momentum in addition to linear one.  
The fact was pointed out by Poynting \cite{Poynting1909} as early as in 1909, 
and  was confirmed experimentally by Beth \cite{Beth1936} in 1936, 
who observed the torque exerted on a birefringent plate as the polarization 
state of the transmitted light was changed. 
The experiment in fact proved the angular momentum associated with spin 
(the spin angular momentum).
Spin degree-of-freedom is now widely used in various research fields. 

Light also can have orbital angular momentum (OAM), but this property
had not well exploited until Allen et al. \cite{Allen1992} discovered
in 1992 that a Laguerre-Gaussian beam can carry OAM in a well-defined manner. 
One remarkable feature of such beams is a characteristic helical phase 
front with phase singularity at the center. 
Its intensity distribution exhibits an annular profile, in particular,
a completely dark spot at the center. 
Based on this property, lights with OAM are often called ``twisted photons''
or ``optical vortices'' in literatures.  
Having such a new degree-of-freedom, twisted photon beams are recognized
as an excellent platform for new science and a myriad of applications 
\cite{Shen2019,Padgett2017,Molina-Terriza2007,Torres-Torner2011}. 
They include fundamental physics concerning interactions between particles
and photons \cite{Babiker2019,GORGONE2019}, 
quantum optics \cite{Mair2001,Leach2009}, micro manipulation of 
particles/materials \cite{He1995,Garces2003}, 
microscopy and imaging \cite{Swartzlander2001,Swartzlander2008,Furhapter2005}, 
optical data transmission \cite{Erhard2018,Gibson2004,Krenn2016}, 
detection of astronomical rotating object \cite{Harwit2003,Tamburini2020},
among others. Now this research field is expanding rapidly.

Many methods are proposed and being actually used to generate light beams
with OAM. 
In the visible region, the common method is use of fork holograms, 
spiral phase plates \cite{X-Wang2018}, lens-based mode 
converters \cite{Beijersbergen1993}, and q-plates \cite{Marrucci2006}.   
In the X-ray region, high-harmonic radiation from a helical undulator 
\cite{Sasaki2008,Bahrdt2013,Kaneyasu2018} and/or coherent emission
from spirally-bunched electrons 
produced by combination of a laser and undulator 
\cite{Hemsing2012,Hemsing2013,Ribic2017} seem very promising methods. 
Less well established is generation of twisted gamma rays. 
Almost all methods proposed so far utilize up-conversion of wavelength by 
Compton backward scattering \cite{Jentschura2011PRD,Jentschura2011EPC,Ivanov-Sebo2011,Stock2015,Petrillo2016,Taira2017,Chen2019}.
In this energy region,
such beams may play an indispensable role in investigating nuclear structure,
spin puzzle of nucleons \cite{Ivanov2011} and phenomena in astrophysics
associated with rotation. 

In this article, we study an alternative method of 
generating high-energy gamma rays with OAM. 
The method utilizes partially-stripped ions (PSIs) as an energy converter; 
accelerated PSIs absorb and re-emit photons with OAM. 
This interesting possibility is suggested in 
Ref.~\cite{Budker2020} in connection with a proposal of 
Gamma Factory~\cite{Bessonov2013,Krasny2019}.
When initial PSIs have Lorentz boost factor of $\gamma$, 
then the energy of photons re-emitted in the backward direction is amplified
by a factor of $4\gamma^2$.
Compared with more traditional backward Compton scattering, 
the process has an advantage of having much bigger fundamental cross section; 
the Rayleigh scattering cross section proportional to square of the resonant
wavelength versus the Thomson scattering cross section proportional
to square of the classical electron radius.

This paper is organized as follows.
In the next section, we present a basic theory of absorption of photons with
OAM by ions. 
The goal in this section is to calculate the absorption rate for
hydrogen-like PSIs.
Calculation is done based on the Dirac theory since high $Z$ ions are used 
as a target.
Then emission rate of photons from excited PSIs is evaluated;  
here we are interested in the emission of multipole photons, 
in particular E2 photons. 
We present the results of our numerical calculations in section 3
and a summary in section \ref{Sec:Summary}. 
Throughout this paper, the natural unit system $c=\hbar=1$ is used.

\section{\label{Sec:TF}Theory and formulas}
In this section, we present the formulas that describe the absorption 
and emission of twisted photon by a hydrogen-like ion.
Before discussing the relevant rates, we summarize
the kinematics of the energy up-conversion. 

We consider the resonant Rayleigh scattering 
$\gamma_i+I\to I^* \to I+\gamma_f$,
where $I$ is a boosted ion in its ground state and $I^*$ represents 
an excited ion. The energy splitting of $I$ and $I^*$ is denoted by
$E_{eg}$.

Assuming the head-on collision, the resonant condition of the excitation
process, $\gamma_i+I\to I^*$,  is expressed as
\begin{equation}
\omega_i=\frac{E_{eg}}{\gamma(1+\beta)}\left(1+\frac{E_{eg}}{2m_I}\right)
        \simeq\frac{E_{eg}}{2\gamma}\,,
\end{equation}
where $\omega_i$ is the angular frequency of the initial photon $\gamma_i$
(in the laboratory frame),
$\beta$ and $\gamma=1/\sqrt{1-\beta^2}$ are the boost factors of 
the initial ion.
An approximate formula for the case of $\gamma\gg 1$%
($\beta\simeq 1$) and $E_{eg}/m_I\ll 1$ is also shown.

The energy of the photon in the emission process, $I^*\to I+\gamma_f$,
is given by
\begin{equation}
\omega_f=\frac{E_{eg}(1+E_{eg}/2m_I)}
              {\gamma+\omega_i/m_I-(\gamma\beta-\omega_i/m_I)\cos\theta_f}
        \simeq\frac{E_{eg}}{\gamma(1-\beta\cos\theta_f)}\,,
\end{equation}
where $\theta_f$ is the polar angle of the emitted photon momentum with
respect to the direction of the ion boost, and an approximate formula
for $\omega_i, E_{eg}\ll m_I$ is also shown. 
In the events of backward scattering, namely $\theta_f=0$,
$\omega_f$ becomes maximal and 
\begin{equation}
\omega_f^\text{max}\simeq 2\gamma E_{eg}\,,
\end{equation}
for $\gamma\gg 1$.
Thus the energy up-conversion factor is 
$\omega_f^\text{max}/\omega_i\simeq 4\gamma^2$.

\begin{figure}
 \centering
 \includegraphics[width=0.462\textwidth]{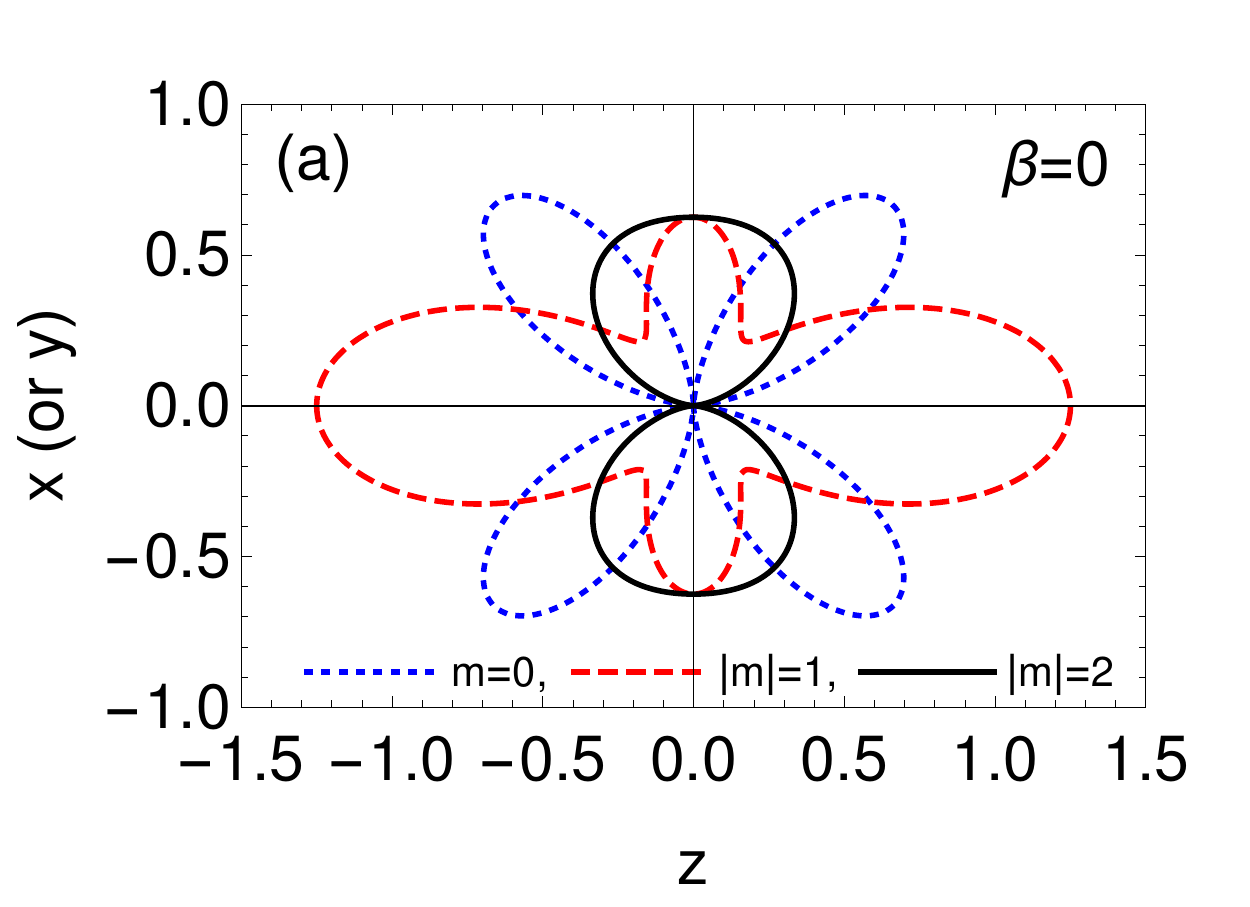}\ 
 \raisebox{0.12em}
   {\includegraphics[width=0.438\textwidth]{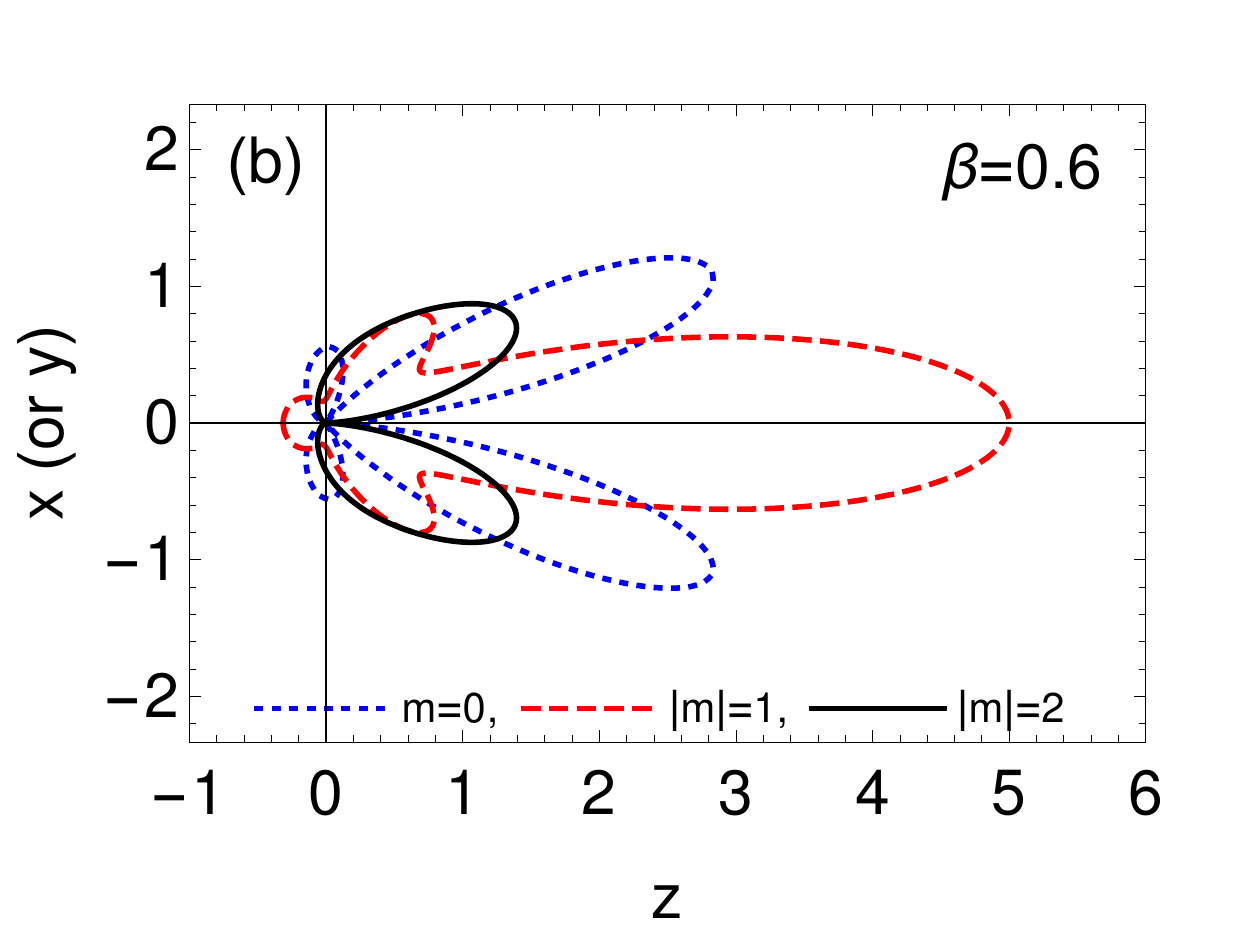}}
 \caption{Angular distributions of the quadrupole radiation
          of $|m|=0,1$ and 2 in dotted blue, dashed red and solid black
          respectively. (a) $\beta=0$ and (b) $\beta=0.6$.}
 \label{Fig:EmissionPattern}
\end{figure}

The emission process is classified by its multipole nature.
Denoting the projection of angular momentum onto the quantization axis
by $m$, a quadrupole or higher multipole radiation of $|m|\neq 1$ 
possesses an OAM and a beam of twisted photons is obtained if the parent 
ions are boosted.
In particular, beams of $|m|\geq 2$ are nontrivial in the sense that
one can access target states of larger magnetic quantum numbers with 
such beams.
It is impossible to make a transition of $|\Delta m|\geq 2$ with
a plane wave in single-photon absorption.%
\footnote{The beam of $m=0$ also possesses an OAM. 
Since it can be obtained in the ordinary electric dipole (E1) 
radiation without using a laser with OAM, we focus on larger $|m|$
in the present work.}
In the numerical illustration in Sec.~\ref{Sec:NR}, 
we consider the deexcitation from the $3d_{5/2}$ state of hydrogen-like 
ions to the ground state, which is dominated by 
the electric quadrupole (E2) radiation.
In Fig.~\ref{Fig:EmissionPattern}, we illustrate
the angular distributions
of the quadrupole radiation in (a) the rest frame of the ion and 
(b) the laboratory frame in which the ion is accelerated as $\beta=0.6$.%
\footnote{In the realistic kinematics of up-conversion in Sec.~\ref{Sec:NR}, 
we employ $\beta\simeq 1$. The rather small $\beta$ here is chosen 
for the visibility of the figure.}
The distribution of $|m|=2$ vanishes on the $z$ axis, which indicates
the phase singularity along the beam axis and exhibits the nature of
twisted photons.

\subsection{Bessel beam}
The Bessel beam is an instance of light beams with OAM. 
In this work, we suppose that a Bessel beam is irradiated
to excite an ion to a state that is able to emit
a twisted photon in its deexcitation process.

A Bessel beam propagating along the $z$ axis is represented by 
the following superposition of plane waves 
\cite{Jentschura2011PRD,Jentschura2011EPC}:
\begin{equation}\label{Eq:FieldSuperposition}
A^\mu_{m_\gamma\kappa_\gamma\lambda}(t_,\bm{r}):=\int
 a_{m_\gamma\kappa_\gamma}(\bm{k}_T)A^\mu_{\bm{k}\lambda}(t_,\bm{r})
\frac{d^2 k_T}{(2\pi)^2}\,,
\end{equation}
where 
\begin{equation}
a_{m_\gamma\kappa_\gamma}(\bm{k}_T):=
 (-i)^{m_\gamma}e^{i m_\gamma\varphi_k}\sqrt{\frac{2\pi}{\kappa_\gamma}}
 \delta(|\bm{k}_T|-\kappa_\gamma)
\end{equation}
is the weight of the superposition for a given photon transverse
wave vector $\bm{k}_T$. 
The plane-wave field of wave vector $\bm{k}$ and helicity $\lambda$ is 
given by
\begin{equation}\label{Eq:PWA}
A^\mu_{\bm{k}\lambda}(t_,\bm{r}):=
 \frac{1}{\sqrt{2\omega}}\epsilon^\mu_\lambda(\bm{k})
                       e^{-i(\omega t-\bm{k}\cdot\bm{r})}\,,\
\omega=|\bm{k}|\,,
\end{equation}
where $\epsilon^\mu_\lambda(\bm{k})$ is the polarization vector.
The wave vector of plane wave is parameterized as 
$\bm{k}=k(\sin\theta_k\cos\varphi_k,\sin\theta_k\sin\varphi_k,\cos\theta_k)$,
so that $\bm{k}\cdot\bm{x}=k_z z+\bm{k}_T\cdot\bm{x}_T$, $k_z=k\cos\theta_k$
and $\bm{k}_T=k\sin\theta_k(\cos\varphi_k,\sin\varphi_k,0)$. 
The polar angle $\theta_k$ is called the pitch angle.
We note that $m_\gamma$ represents the eigenvalue of $J_z$, the $z$ 
component of the total angular momentum, and
$\kappa_\gamma=|\bm{k}_T|=|\bm{k}|\sin\theta_k$.

When a Bessel beam is irradiated on an ion, 
the target ion is not always on the axis of the Bessel beam.
We denote the impact parameter of the Bessel beam by $\bm{b}$ taking
the position of the ionic nucleus as the origin of the coordinate
system.
Then, the electromagnetic field at the electronic position $\bm{r}$ is
given by Eq.~\eqref{Eq:FieldSuperposition} with 
\cite{Matula2013JPB}
\begin{equation}
A^\mu_{\bm{k}\lambda}(t_,\bm{r}):=
 \frac{1}{\sqrt{2\omega}}\epsilon^\mu_\lambda(\bm{k})
 e^{-i(\omega t-\bm{k}\cdot\bm{r}-\bm{k}_T\cdot\bm{b})}\,.
\end{equation}

\subsection{Absorption of twisted photons by hydrogen-like heavy ions}
In the following, we evaluate the absorption cross section 
in the rest frame of the ion.
In this frame the resonant energy of the incident photon is $E_{eg}$, 
which is of order keV for heavy ions.

The interaction of the electron with the radiation field in 
the Dirac theory is given by 
\begin{equation}
H_I=e\bm{\alpha}\cdot\bm{A}\,,
\end{equation}
where $e=|e|$ is the unit charge and $\bm{\alpha}$ represents the Dirac
matrices. 
The absorption amplitude is given by a matrix element of
$V=H_Ie^{i\omega t}$.

\subsubsection{Amplitude superposition}
It is convenient to express the twisted photon amplitude
as a superposition of plane wave amplitudes like
the twisted photon field itself.
Then, the absorption amplitude is written as \cite{Scholtz-Marggraf2014PRA}
\begin{equation}\label{Eq:AmpSP}
M_{fi}^\text{(tw)}=\langle f|V^\text{(tw)}|i\rangle
=\int\frac{d^2 k_T}{(2\pi)^2}
  a_{m_\gamma\kappa_\gamma}(\bm{k}_T)e^{-i\bm{k}_T\cdot\bm{b}}
  M_{fi}^\text{(pl)}(\theta_k,\varphi_k)
\end{equation}
where
$M_{fi}^\text{(pl)}(\theta_k,\varphi_k):=\langle f|V^\text{(pl)}|i\rangle$
is the plane wave amplitude in which the direction of the photon wave
vector is specified by $\theta_k$ and $\varphi_k$ with respect to 
the ionic spin quantization axis (taken to be the $z$ axis).
Such an inclined plane wave amplitude is expressed as
\begin{equation}\label{Eq:InclinedM}
M_{fi}^\text{(pl)}(\theta_k,\varphi_k)
=e^{-i(m_f-m_i)\varphi_k}\sum_{m'_f m'_i}
  d_{m_fm'_f}^{j_f}(\theta_k)d_{m_im'_i}^{j_i}(\theta_k)M_{m'_fm'_i}\,,
\end{equation}
where $d_{mm'}^j(\theta)$ is Wigner's d-function and
$M_{m'_f m'_i}$ represents the ordinary plane wave amplitude 
of $\theta_k=\varphi_k=0$.
Substituting Eq.~\eqref{Eq:InclinedM} into Eq.~\eqref{Eq:AmpSP}, 
one obtains
\begin{align}
M_{fi}^\text{(tw)}
&=(-i)^{2m_\gamma+m_i-m_f}e^{i(m_\gamma+m_i-m_f)\varphi_b}
 \sqrt{\frac{\kappa_\gamma}{2\pi}}J_{m_\gamma+m_i-m_f}(\kappa_\gamma b)
 M_{fi}^\text{(pl)}(\theta_k,0)\nonumber\\
&=:\mathcal{M}_{m_fm_i}(\bm{b})\,,\label{Eq:AmpTWb}
\end{align}
where $\varphi_b$ denotes the azimuthal angle of $\bm{b}$.

\subsubsection{Plane wave amplitude in the Dirac theory}
Here, we consider hydrogen-like ions in the Dirac theory.
(See e.g.~Ref.~\cite{Sakurai1967a}.)
The wave function is given by
\begin{equation}\label{Eq:DiracWF}
\psi_{n\kappa m}(\bm{r})
=\begin{pmatrix}
  \displaystyle{\frac{G_{n\kappa}(r)}{r}}\Omega_{jm}^{\ell_A}(\theta,\varphi)
  \\[2ex]
  i\displaystyle{\frac{F_{n\kappa}(r)}{r}}\Omega_{jm}^{\ell_B}(\theta,\varphi)
\end{pmatrix}\,,
\end{equation}
where $\kappa=\mp(j+1/2)$, $\ell_A=j\mp 1/2$, $\ell_B=j\pm 1/2$ and
$\Omega_{jm}^\ell(\theta,\varphi)$ denotes the spinor spherical harmonics.
The wave function is normalized as 
$\int d^3r\,|\psi_{n\kappa m}(\bm{r})|^2
 =\int dr\,\{G_{n\kappa}^2(r)+F_{n\kappa}^2(r)\}=1$.
The electronic state of an ion may be denoted as $|n,\kappa,m\rangle$.

The plane wave matrix element in the right-hand side of
Eq.~\eqref{Eq:InclinedM} is expressed as
\begin{equation}
M_{m_fm_i}=\langle n_f,\kappa_f,m_f|V^\text{(pl)}|n_i,\kappa_i,m_i\rangle\,.
\end{equation}
Using the Dirac wave function in Eq.~\eqref{Eq:DiracWF} and 
the plane wave field in Eq.~\eqref{Eq:PWA}, we find that
\begin{align}
M_{m_fm_i}
=&-ie\sqrt{\frac{3}{\omega}}(-1)^{j_i+m_f}\sqrt{(2j_f+1)(2j_i+1)}\,
 \delta_{m_f,m_i+\lambda}\nonumber\\
 &\times\sum_{j,\ell}(-1)^ji^\ell\sqrt{2\ell+1}\,
               C_{j_f-m_f,j_im_i}^{j-m_f+m_i}
               C_{\ell-m_f+m_i+\lambda,1-\lambda}^{j-m_f+m_i}\nonumber\\
 &\phantom{\times\sum}
  \left[(-1)^{\ell_{Af}}\sqrt{(2\ell_{Af}+1)(2\ell_{Bi}+1)}\,
        C_{\ell_{Af}0,\ell_{Bi}0}^{\ell 0}
        \begin{Bmatrix}
         \ell_{Af} & j_f & 1/2\\
         \ell_{Bi} & j_i & 1/2\\
         \ell      & j   & 1  
        \end{Bmatrix}\mathcal{I}_{GF}^\ell\right.\nonumber\\
 &\phantom{\times\sum}
  \left.-(-1)^{\ell_{Bf}}\sqrt{(2\ell_{Bf}+1)(2\ell_{Ai}+1)}\,
        C_{\ell_{Bf}0,\ell_{Ai}0}^{\ell 0}
        \begin{Bmatrix}
         \ell_{Bf} & j_f & 1/2\\
         \ell_{Ai} & j_i & 1/2\\
         \ell      & j   & 1  
        \end{Bmatrix}\mathcal{I}_{FG}^\ell\right]\,,
\end{align}
where $C_{j_1m_1,j_2m_2}^{j_3m_3}$ is the Clebsch-Gordan coefficient and
\begin{equation}
 \begin{Bmatrix}
  a & b & c\\
  d & e & f\\
  g & h & i  
 \end{Bmatrix}
\end{equation}
represents the $9j$ symbol, e.g.~\cite{VMK1988a}.
The radial integrals are defined by
\begin{equation}
\mathcal{I}_{GF}^\ell:=\int dr\,j_\ell(kr)G_f(r)F_i(r)\,,\ 
\mathcal{I}_{FG}^\ell:=\int dr\,j_\ell(kr)F_f(r)G_i(r)\,,
\end{equation}
where $j_\ell(kr)$ denotes the spherical Bessel function and
$k=|\bm{k}|$ is the wave number of the plane wave photon.

\subsubsection{Absorption cross section}
The absorption rate of a twisted photon is proportional to
the squared amplitude in Eq.~\eqref{Eq:AmpTWb}
\begin{equation}
|\mathcal{M}_{m_fm_i}(\bm{b})|^2
=(\kappa_\gamma/2\pi)|M^\text{(pl)}_{m_fm_i}(\theta_k,0)|^2
 J^2_{m_\gamma+m_i-m_f}(\kappa_\gamma b)\,.
\end{equation}
In experiments where the ion beam is not sufficiently collimated 
(as virtually all experiments we envisage), it is legitimate to
average over the impact parameter $\bm{b}$. 
Provided that the (effective) beam radius $R$ is 
large enough as 
$\kappa_\gamma R\gg1$, (fat beam approximation), one obtains 
\begin{equation}
\frac{1}{\pi R^2}\int|\mathcal{M}_{m_fm_i}(\bm{b})|^2 d^2b
\simeq \frac{1}{\pi^2 R}|M^\text{(pl)}_{m_fm_i}(\theta_k,0)|^2\,,
\end{equation}
where we have used the asymptotic form of the Bessel function,
$J_m(z)\simeq\sqrt{2/\pi z}\,\cos(z-\pi/4-m\pi/2)$.
We note that the $m_\gamma$ dependence disappears in this approximation.

The photon number flux of the Bessel beam is given by
$f(R)=\cos\theta_k/\pi^2 R$ in the fat beam 
approximation \cite{Jentschura2011EPC}.
With this flux, we obtain the absorption cross section for 
a given set of initial and final magnetic quantum numbers as
\begin{equation}\label{Eq:CS}
\sigma(\omega)=\frac{2}{\cos\theta_k}
               \frac{\Gamma_f/2}{(\omega+E_i-E_f)^2+\Gamma_f^2/4}
               |M^\text{(pl)}_{m_fm_i}(\theta_k,0)|^2\,,
\end{equation}
where $E_{i(f)}$ is the energy of the initial (final) state and
$\Gamma_f$ denotes the natural width of the final state.
\footnote{Here we ignore the width of excitation laser and other
broadening effects for simplicity.}

\subsection{Emission of twisted photons by hydrogen-like heavy ions} 
The photons emitted in multipole radiations of $j\geq 2$ are 
twisted in the sense that they have nonzero orbital angular momentum.
Accordingly, we consider the multipole radiations of hydrogen-like
heavy ions.

The interaction hamiltonian $H_I=-e\gamma_\mu A^\mu$ in which 
the electromagnetic field is one of the multipole fields introduced below
describes the multipole emission. 
The emission rate of $|i\rangle\to|f\rangle+\gamma$ is given by
\begin{equation}\label{Eq:EmissionRate}
dw=2\pi|V_{fi}|^2\delta(E_i-E_f-\omega)d\omega\,,\ 
w=2\pi|V_{fi}|^2\,,
\end{equation}
where $V_{fi}=\langle f|H_I e^{-i\omega t}|i\rangle$ is the emission
matrix element.
We work in the rest frame of the ion as in the evaluation of 
the absorption cross section.

We do not employ the long wavelength approximation.
The scale of transition wavelengths is $1/(Z\alpha)^2 m_e$,
and the size of an ion is $\sim 1/Z\alpha m_e$.
In the case $Z\alpha\ll 1$, the long wavelength approximation is legitimate.
For heavy ions of $Z\alpha=O(1)$, 
the long wavelength approximation is questionable. 

\subsubsection{Multipole fields}
In the Coulomb gauge, the multipole vector potentials are written in
terms of the vector spherical harmonics, $\bm{Y}_{jm}^\ell(\bm{n})$.
We note that
\begin{align}
&\hat{\bm{J}}^2\bm{Y}_{jm}^\ell=j(j+1)\bm{Y}_{jm}^\ell\,,\ 
\hat{J}_z\bm{Y}_{jm}^\ell=m\bm{Y}_{jm}^\ell\,,\\
&\hat{\bm{L}}^2\bm{Y}_{jm}^\ell=\ell(\ell+1)\bm{Y}_{jm}^\ell\,,\ 
\hat{\bm{S}}^2\bm{Y}_{jm}^\ell=2\bm{Y}_{jm}^\ell\,.
\end{align}
The vector spherical harmonics may be expressed as
$\bm{Y}_{jm}^\ell(\bm{n})
=\sum_{m,\sigma}C^{jm}_{\ell m, 1 \sigma}Y_{\ell m}(\bm{n})\bm{e}_\sigma$,
where $\bm{e}_\sigma$ denotes the covariant spherical basis vectors,
$\bm{e}_{\pm 1}=\mp(\bm{e}_x\pm i\bm{e}_y)/\sqrt{2}$ and 
$\bm{e}_0=\bm{e}_z$ \cite{VMK1988a}.

The electric multipole field of angular frequency $\omega$ and wave vector 
$\bm{k}$ is given by%
\footnote{The Gauss units in Ref.~\cite{LLQED} and 
the Heaviside-Lorentz rationalized units employed here are related by 
$\alpha=e_\text{G}^2=e_\text{HL}^2/(4\pi)$ and 
$e_\text{G}A_\text{G}=e_\text{HL}A_\text{HL}$.}
\begin{equation}\label{Eq:EMF}
\bm{A}_{\omega jm}(\bm{k})=2(\pi/\omega)^{3/2}\delta(|\bm{k}|-\omega)
                           \bm{Y}^{(e)}_{jm}(\hat{\bm{k}})\,,
\end{equation}
where
\begin{equation}\label{Eq:YE}
\bm{Y}^{(e)}_{jm}(\hat{\bm{k}})
:=\frac{1}{\sqrt{j(j+1)}}\bm{\nabla}_{\hat{\bm{k}}}Y_{jm}(\hat{\bm{k}})
=\sqrt{\frac{j+1}{2j+1}}\bm{Y}_{jm}^{j-1}
 +\sqrt{\frac{j}{2j+1}}\bm{Y}_{jm}^{j+1}\,.
\end{equation}
The magnetic multipole field is written as
\begin{equation}\label{Eq:MMF}
\bm{A}_{\omega jm}(\bm{k})=2(\pi/\omega)^{3/2}\delta(|\bm{k}|-\omega)
                           \bm{Y}^{(m)}_{jm}(\hat{\bm{k}})\,,
\end{equation}
where
\begin{equation}
\bm{Y}^{(m)}_{jm}(\hat{\bm{k}})
:=\hat{\bm{k}}\times\bm{Y}^{(e)}_{jm}(\hat{\bm{k}})
=i\bm{Y}_{jm}^j\,.
\end{equation}
We note that
\begin{equation}
\int d\Omega_{\hat{\bm{k}}}
 \bm{Y}^{(e)*}_{jm}(\hat{\bm{k}})\cdot\bm{Y}^{(e)}_{j'm'}(\hat{\bm{k}})
=\int d\Omega_{\hat{\bm{k}}}
 \bm{Y}^{(m)*}_{jm}(\hat{\bm{k}})\cdot\bm{Y}^{(m)}_{j'm'}(\hat{\bm{k}})
=\delta_{jj'}\delta_{mm'}\,.
\end{equation}

The parities of the electric and magnetic multipole fields are opposite.
We assign $(-1)^j$ to the electric multipole field in Eq.~\eqref{Eq:EMF}
and $(-1)^{j+1}$ to the magnetic multipole field in Eq.~\eqref{Eq:MMF}.
For instance, the electric (magnetic) dipole field is parity odd (even).

In the coordinate space, the field is given by%
\footnote{The time dependence of $e^{-i\omega t}$ is implicit.}
\begin{equation}
\bm{A}_{\omega jm}(\bm{r})
=\int\frac{dk^3}{(2\pi)^3}\bm{A}_{\omega jm}(\bm{k})e^{i\bm{k}\cdot\bm{r}}\,.
\end{equation}

\subsubsection{Electric multipole radiations in the Dirac theory}
Since the electric multipole field contains two components of orbital
angular momentum, $\ell=j\pm 1$, as seen in Eq.~\eqref{Eq:YE},
we rearrange them by a gauge transformation \cite{LLQED}. 
A gauge transformation results in the following vector and scalar
potentials:
\begin{align}
&\bm{A}_{\omega jm}(\bm{k})
 =2(\pi/\omega)^{3/2}\delta(|\bm{k}|-\omega)
  \left\{\bm{Y}^{(e)}_{jm}(\hat{\bm{k}})+C\hat{\bm{k}}Y_{jm}(\hat{\bm{k}})
  \right\}\,,\\
&\phi_{\omega jm}(\bm{k})
 =2(\pi/\omega)^{3/2}\delta(|\bm{k}|-\omega)CY_{jm}(\hat{\bm{k}})\,,
\end{align}
where $C$ is a gauge parameter. 
Choosing $C=-\sqrt{(j+1)/j}$, we obtain
\begin{align}
&\bm{A}_{\omega jm}^{(e)}(\bm{k})
 =\sqrt{(2j+1)/j}\,2(\pi/\omega)^{3/2}\delta(|\bm{k}|-\omega)
  \bm{Y}_{jm}^{j+1}(\hat{\bm{k}})\,,\label{Eq:EMA}\\
&\phi_{\omega jm}^{(e)}(\bm{k})
 =-\sqrt{(j+1)/j}\,2(\pi/\omega)^{3/2}\delta(|\bm{k}|-\omega)
  Y_{jm}(\hat{\bm{k}})\,. \label{Eq:EMphi}
\end{align}
In the long wavelength approximation, $\phi_{\omega jm}^{(e)}$ 
dominantly contributes and $\bm{A}_{\omega jm}^{(e)}(\bm{k})$ does
subdominantly. 
As we mentioned above, we consider both contributions for heavy ions.

The contribution of the scalar potential in Eq.~\eqref{Eq:EMphi}
to the emission matrix element is given by
\begin{align}
V_{fi}^{(\phi)}&=-e\int d^3r\psi_f^\dagger(\bm{r})\psi_i(\bm{r})
                 \int\frac{d^3k}{(2\pi)^3}
                  \phi_{\omega jm}^{(e)*}(\bm{k})e^{-i\bm{k}\cdot\bm{r}}\\
 &=e\sqrt{\frac{j+1}{j}}\frac{\sqrt{\omega}}{4\pi^{3/2}}
   \int d^3r\psi_f^\dagger(\bm{r})\psi_i(\bm{r})
   \int d\Omega_k e^{-i\bm{k}\cdot\bm{r}}Y_{jm}^*(\hat{\bm{k}})\,,
\end{align}
where $\psi_{i(f)}(\bm{r})$ represents the initial (final) wave function
given in Eq.~\eqref{Eq:DiracWF}.
The angular integral over $\Omega_k$ is performed using
\begin{equation}
e^{i\bm{k}\cdot\bm{r}}
=4\pi\sum_{\ell=0}^{\infty}\sum_{m=-\ell}^{m=\ell}
  i^\ell j_\ell(kr)Y^*_{\ell m}(\hat{\bm{k}})Y_{\ell m}(\hat{\bm{r}})\,,
\end{equation}
and one obtains
\begin{equation}
V_{fi}^{(\phi)}
=e(-i)^j\sqrt{\frac{j+1}{j}}\sqrt{\frac{\omega}{\pi}}
 \int d^3r\psi_f^\dagger(\bm{r})\psi_i(\bm{r})j_j(kr)Y_{jm}^*(\hat{\bm{r}})\,.
\end{equation}
Substituting the Dirac wave function in Eq.~\eqref{Eq:DiracWF} and
integrating over the remaining angular variables, we obtain
\begin{align}
V_{fi}^{(\phi)}=&\frac{e}{2\pi}\sqrt{\omega}\,i^j(-1)^{j_f+m_f+j_i+1/2}\,
                 \sqrt{\frac{(j+1)(2j_f+1)(2j_i+1)}{j(2j+1)}}\,
                 C^{j-m_f+m_i}_{j_f-m_f,j_im_i}\nonumber\\
&\times\left[\begin{Bmatrix}
              \ell_{Af} & \ell_{Ai} & j\\
              j_i       & j_f       & 1/2
             \end{Bmatrix}
             \sqrt{(2\ell_{Af}+1)(2\ell_{Ai}+1)}\,
             C^{j0}_{\ell_{Af}0,\ell_{Ai}0}\,\mathcal{I}^j_{GG}\right.
             \nonumber\\
            &\quad\left.
            +\begin{Bmatrix}
              \ell_{Bf} & \ell_{Bi} & j\\
              j_i       & j_f       & 1/2
             \end{Bmatrix}
             \sqrt{(2\ell_{Bf}+1)(2\ell_{Bi}+1)}\,
             C^{j0}_{\ell_{Bf}0,\ell_{Bi}0}\,\mathcal{I}^j_{FF}
       \right]\,,\label{Eq:Vphi}
\end{align}
where 
\begin{equation}
 \begin{Bmatrix}
  a & b & c\\
  d & e & f
 \end{Bmatrix}
\end{equation}
represents the $6j$ symbol and the radial integrals are given by
\begin{equation}
\mathcal{I}_{GG}^j:=\int dr\,j_\ell(kr)G_f(r)G_i(r)\,,\ 
\mathcal{I}_{FF}^j:=\int dr\,j_\ell(kr)F_f(r)F_i(r)\,.
\end{equation}

As for the contribution of the vector potential in Eq.~\eqref{Eq:EMA},
the $\bm{k}$ integration results in
\begin{equation}
V_{fi}^{(A)}
=e(-i)^{j+1}\sqrt{\frac{2j+1}{j}}\sqrt{\frac{\omega}{\pi}}
 \int d^3r j_{j+1}(kr)\psi_f^\dagger(\bm{r})\bm{\alpha}\psi_i(\bm{r})\cdot
  \bm{Y}_{jm}^{j+1*}(\hat{\bm{r}})\,.
\end{equation}
Integrating over $\bm{r}$, we find
\begin{align}
V_{fi}^{(A)}=&\frac{e}{2\pi}\sqrt{\omega}\,i^j(-1)^{j_i+m_f}\,
                 \sqrt{\frac{6(2j+1)(2j_f+1)(2j_i+1)}{j}}\,
                 C^{j-m_f+m_i}_{j_f-m_f,j_im_i}\nonumber\\
&\times\left[\begin{Bmatrix}
              \ell_{Af} & j_f & 1/2\\
              \ell_{Bi} & j_i & 1/2\\
              j+1       & j   & 1
             \end{Bmatrix}
             (-1)^{\ell_{Af}}\sqrt{(2\ell_{Af}+1)(2\ell_{Bi}+1)}\,
             C^{j+10}_{\ell_{Af}0,\ell_{Bi}0}\,\mathcal{I}^{j+1}_{GF}\right.
             \nonumber\\
            &\quad\left.
            -\begin{Bmatrix}
              \ell_{Bf} & j_f & 1/2\\
              \ell_{Ai} & j_i & 1/2\\
              j+1       & j   & 1
             \end{Bmatrix}
             (-1)^{\ell_{Bf}}\sqrt{(2\ell_{Bf}+1)(2\ell_{Ai}+1)}\,
             C^{j+10}_{\ell_{Bf}0,\ell_{Ai}0}\,\mathcal{I}^{j+1}_{FG}
       \right]\,.\label{Eq:VA}
\end{align}
The total amplitude is $V_{fi}=V_{fi}^{(\phi)}+V_{fi}^{(A)}$.

\subsubsection{Magnetic multipole radiations in the Dirac theory}
The magnetic multipole radiation is described by the vector potential
in Eq.~\eqref{Eq:MMF}. 
The emission matrix element is evaluated in the similar manner as
$V_{fi}^{(A)}$ above and we obtain
\begin{align}
V_{fi}=&\frac{e}{2\pi}\sqrt{\omega}\,i^j(-1)^{j_i+m_f}\,
        \sqrt{6(2j_f+1)(2j_i+1)}\,C^{j-m_f+m_i}_{j_f-m_f,j_im_i}
        \nonumber\\
&\times\left[\begin{Bmatrix}
              \ell_{Af} & j_f & 1/2\\
              \ell_{Bi} & j_i & 1/2\\
              j         & j   & 1
             \end{Bmatrix}
             (-1)^{\ell_{Af}}\sqrt{(2\ell_{Af}+1)(2\ell_{Bi}+1)}\,
             C^{j0}_{\ell_{Af}0,\ell_{Bi}0}\,\mathcal{I}^{j}_{GF}\right.
             \nonumber\\
            &\quad\left.
            -\begin{Bmatrix}
              \ell_{Bf} & j_f & 1/2\\
              \ell_{Ai} & j_i & 1/2\\
              j         & j   & 1
             \end{Bmatrix}
             (-1)^{\ell_{Bf}}\sqrt{(2\ell_{Bf}+1)(2\ell_{Ai}+1)}\,
             C^{j0}_{\ell_{Bf}0,\ell_{Ai}0}\,\mathcal{I}^{j}_{FG}
       \right]\,.
\end{align}

\section{\label{Sec:NR}Numerical results}
As an illustration, we consider the excitation of H-like ions
from the ground state $1\text{s}_{1/2}$ to the excited state 
of $3\text{d}_{5/2}$ by the Bessel beam and
the successive deexcitation back to the ground state by the
E2 emission.
If the magnetic quantum number of the ground state differs from
that of the exited state by two (or larger), this process is not
possible with the plane-wave beam nor the dipole emission.

For instance, one may consider $2\text{p}_{3/2}$ instead of 
$3\text{d}_{5/2}$. 
The excitation to the state of $|m|=3/2$ is possible with 
the plane-wave beam. This is an E1 transition and its rate is significant.
However, the subsequent deexcitation from $2\text{p}_{3/2}(m=\pm 3/2)$
to $1\text{s}_{1/2}(m=\mp 1/2)$, which generates a twisted gamma ray
of $|m|=2$, is the magnetic quadrupole (M2) transition.
It turns out that, even for Pb, this M2 rate is suppressed by a factor
of $\sim 10^3$ compared to the dominant E1 emission of an untwisted photon.
This E1 photon becomes a serious background because its energy is
the same as the M2 photon.%
\footnote{The level splitting due to the Zeeman effect is much smaller
than the $2\text{p}_{3/2}$ natural width, not sufficient to separate
the M2 photons from the E1 photons, unlike the case discussed in 
Sec.~\ref{Sec:Summary}.}

\subsection{\label{Subsec:ExR}Excitation rate with twisted photons}
\begin{figure}
 \centering
 \raisebox{-0.14em}
  {\includegraphics[width=0.446\textwidth]{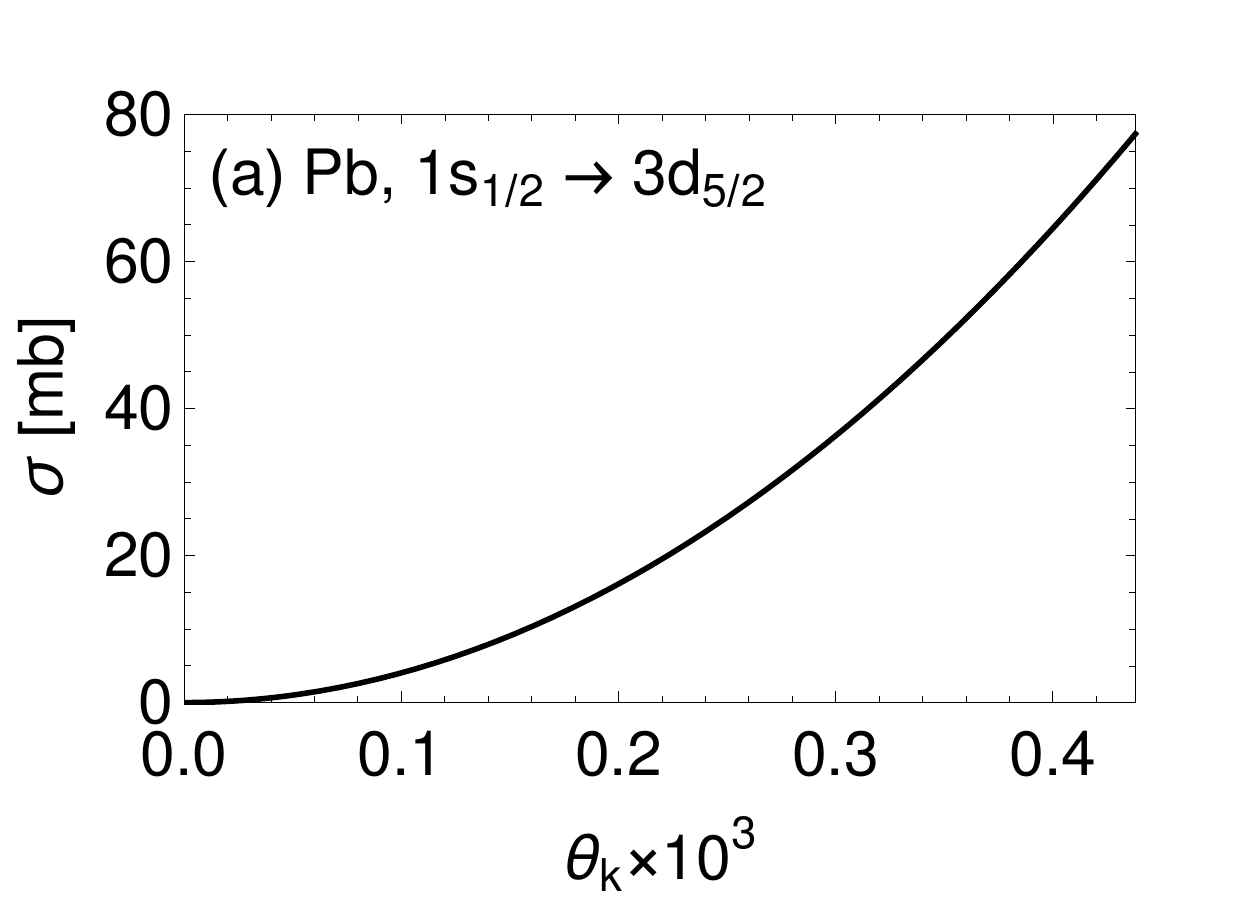}}\ 
 \includegraphics[width=0.454\textwidth]{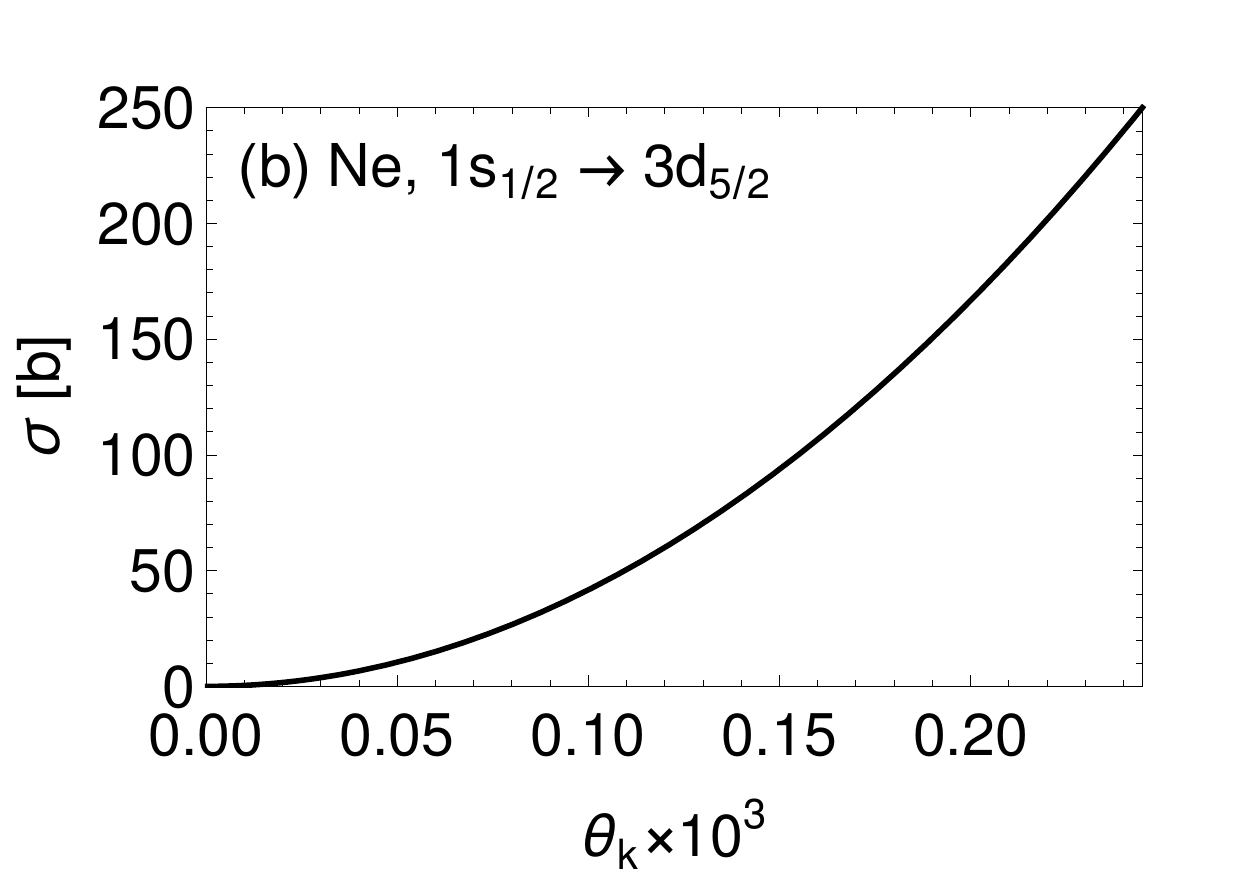}
 \caption{Absorption cross section of 
          $1\text{s}_{1/2,\,m=1/2}\to 3\text{d}_{5/2,\,m=5/2}$.
          The photon helicity is $\lambda=1$.
          (a) The target ion is Pb with $\omega_i=10$ eV.
          (b) Ne with $\omega_i=1$ eV.
          We note the difference in the units of the vertical axes.} 

 \label{Fig:thetaCS}
\end{figure}

The resonant excitation rate is described by the absorption cross
section in Eq.~\eqref{Eq:CS} with 
$\omega=E_{eg}=E(3\text{d}_{5/2})-E(1\text{s}_{1/2})$ and
presented in Fig.~\ref{Fig:thetaCS} as a function of the pitch angle
$\theta_k$.  
We choose $m_i=1/2$, $m_f=5/2$ and $\lambda=1$.

The target ion is H-like Pb ($Z=82$) in Fig.~\ref{Fig:thetaCS}(a). 
The level splitting is $E_{eg}=91.3$ keV.
The energy of the excitation laser is taken as $\omega_i=10$ eV, and
this implies $\gamma=4.57\times 10^3$ and $\omega_f^\text{max}=0.834$ GeV.
The width of the excited state is dominated by the major E1 rate,
$\Gamma_f\simeq\Gamma(3\text{d}_{5/2}\to 2\text{p}_{3/2})=1.89$ eV.

In Fig.~\ref{Fig:thetaCS}(b), the ion is H-like Ne ($Z=10$) and we take
$\omega_i=1$ eV. 
The relevant parameters in this case are $\gamma=606$, 
$\omega_f^\text{max}=1.47$ MeV and 
$\Gamma_f\simeq 4.26\times 10^{-4}$ eV.
The much larger cross section of Ne than Pb is due to the narrow
width of the $3\text{d}_{5/2}$ state.

The pitch angle $\theta_k$ can be $O(1)$ in the laboratory frame.
While, in the rest frame of the ion, it is $O(1/\gamma)$ because
the transverse momentum $\bm{k}_T$ of twisted photons is invariant 
under the Lorentz boost. 
We have chosen the ranges of the horizontal axis in Fig.~\ref{Fig:thetaCS}
following this observation and the difference of $\omega_i$'s for
the two species of ion.
The absorption cross section is proportional to $\theta_k^2$ for 
$|\theta_k|\ll 1$ as seen in Fig.~\ref{Fig:thetaCS}.
The case of $\theta_k=0$ corresponds to a plane wave, 
and the cross section vanishes since the process of $|m_f-m_i|>1$
is impossible with the plane wave as mentioned above.

\subsection{\label{Subsec:EmR}Emission rate of twisted photons}
\begin{figure}
 \centering
 \raisebox{0.05em}
  {\includegraphics[width=0.4456\textwidth]{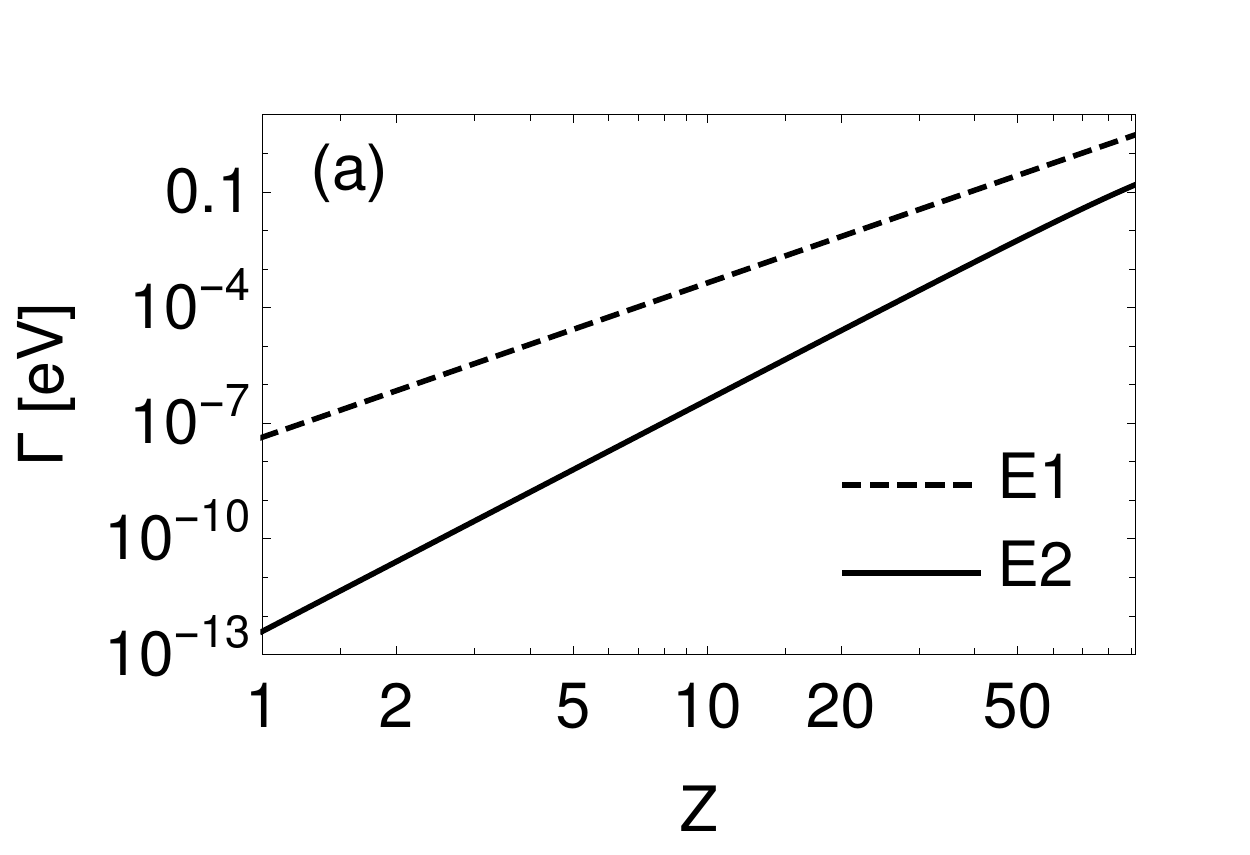}}\ 
 \includegraphics[width=0.4544\textwidth]{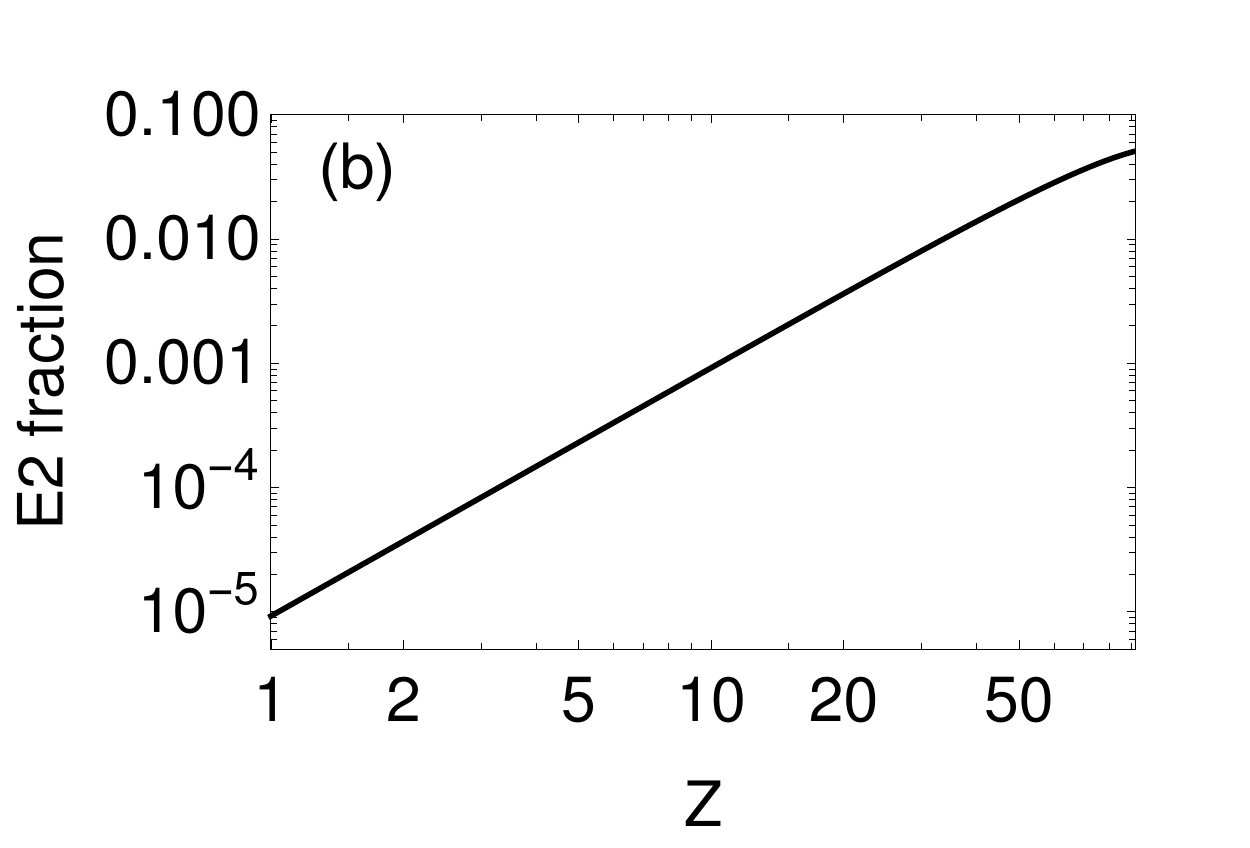}\ 
 \caption{(a) E2 rate compared to E1. (b) Ratio of E2 to E1.}
 \label{Fig:ZE1E2}
\end{figure}

The E2 transition rate from $3\text{d}_{5/2}$ to $1\text{s}_{1/2}$
is given by Eqs.~\eqref{Eq:EmissionRate}, \eqref{Eq:Vphi} and \eqref{Eq:VA}
with $j=2$. 
In Fig.~\ref{Fig:ZE1E2}(a), 
we present the E2 transition rate of H-like ion as a function of $Z$. 
The dominant E1 rate from $3\text{d}_{5/2}$ to $2\text{p}_{3/2}$ is
also shown for comparison.
The ratio of the E2 rate to the E1 rate, which gives an approximate
branching fraction to emit an E2 photon, is given in Fig.~\ref{Fig:ZE1E2}(b).
We observe that the E2 rate depends on $Z$ as $Z^6$ in good precision,
while the E1 is approximately proportional to $Z^4$,
so that heavier ions exhibit larger branching fractions
of emitting a twisted photon.

For H-like Pb, we find 
$\Gamma(3\text{d}_{5/2}\to 1\text{s}_{1/2})=8.49\times 10^{-2}$ eV,
so that the branching fraction of the E2 transition is
$\text{Br}(3\text{d}_{5/2}\to 1\text{s}_{1/2})\simeq 4.5\times 10^{-2}$.
As for H-like Ne, 
$\Gamma(3\text{d}_{5/2}\to 1\text{s}_{1/2})=3.89\times 10^{-7}$ eV,
and $\text{Br}(3\text{d}_{5/2}\to 1\text{s}_{1/2})\simeq 9.1\times 10^{-4}$.

\subsection{Characteristics of photons generated by the method}
In the excitation process in Sec.~\ref{Subsec:ExR}, the final state
of $m_f=3/2$, in addition to $m_f=5/2$,
is also possible with the same Bessel beam.
Then, the $3\text{d}_{5/2}(m=3/2)$ state deexcites to the ground state 
($1\text{s}_{1/2}$) by the E2 emission of $m=1$.
Its emission pattern is shown in Fig.~\ref{Fig:EmissionPattern}(b) 
(the red dashed line) and exhibits no phase singularity on the $z$ axis.
So that, the photon in this process is not twisted and could be
a source of backgrounds.

It turns out that the $m_f=3/2$ process dominates the absorption
cross section as 
$\sigma_{m_f=5/2}/\sigma_{m_f=3/2}\simeq 6\times 10^{-8}(3\times 10^{-6})$
for H-like Pb (Ne) in the case of $\theta=1/\gamma$.
This is because the process of $|m_f-m_i|=1$ is possible with the plane
wave and the cross section does not vanish even if $\theta_k=0$.
We note that the above ratios are $O(\theta_k^2)=O(1/\gamma^2)$.

\section{\label{Sec:Summary}Summary}
We first point out two important aspects of the method 
we proposed in this article: 
one is its final flux of gamma rays with OAM and the other is associated 
backgrounds. 
As shown in Fig.~\ref{Fig:thetaCS}, the absorption cross sections are 
in a range of a few tens of mb (for $Z=82$) to b (for $Z=10$). 
The achievable flux depends heavily upon actual experimental configurations, 
in particular accelerators and incident lasers. 
Considering current technologies, we expect reasonable flux useful to 
a variety of physics.%
\footnote{%
For example, we expect an order of $10^4$ excitations per second to $m_f = 5/2$ 
in the case of Pb 
with the following parameters; 
$\sigma=$ 40 mb, $1 \times 10^9$ H-like ions/bunch with a fractional energy
spread of $2 \times 10^{-4}$, a 100 W laser focusing on a spot size of
1 mm$^2$, a 10-m-long interaction section,
and 20 MHz bunch repetition rate \cite{Krasny2019}.}

As to the backgrounds, we expect two major backgrounds in this method 
as discussed in the previous section: 
one is due to the process from d-states to p-states and the other from 
$m_f=3/2$ of the d-states to s-states. 
The former type of the backgrounds has a different energy from that of
the signal, i.e. the gamma rays with OAM. 
Whether they are tolerable or not depends upon details of particular
experiments, for example detectors employed or reactions in interests 
may or may not be sensitive to such backgrounds.  
In general, however, it is desirable to reduce them; from this view point, 
we believe it important to consider other ion types than the hydrogen-like
ions. 
On the other hand, the second type of the backgrounds has the same energy 
with the signal. 
In this case, it is highly recommended to eliminate or reduce these 
backgrounds. 
One way to eliminate such backgrounds is to introduce a transverse magnetic
field $\bm{B}_T$ in the production region.  
The Zeeman effect splits each magnetic quantum number $m_f$, 
one of which (i.e. $m_f=5/2$) can be selected by choosing
the frequency of the excitation photons. 
Note that $\bm{B}_T$ is amplified by the Lorentz boost factor $\gamma$ when 
seen by the ions. 
Also note that it is necessary to rotate the spin from transverse to 
longitudinal. 
Designing actual spin rotation systems
requires more detailed studies, which is underway currently. 

In summary, we have studied an alternative method of 
generating gamma rays with OAM.
It exploits accelerated partially-stripped ions as an energy upconverter.  
Relativistic calculations are performed to calculate the excitation cross 
section and deexcitation rate for hydrogen-like ions, and 
their properties including flux and possible backgrounds are discussed.

\section*{Acknowledgments}
The work of MT is supported in part by JSPS KAKENHI Grant Numbers 
JP 16H03993 and 18K03621.
The work of NS is supported in part by JSPS KAKENHI Grant Number
JP 16H02136.

\end{document}